# Some Observations on the Performance of the Most Recent Exchange-Correlation Functionals for the Large and Chemically Diverse GMTKN55 Benchmark

Golokesh Santra[1] and Jan M.L. Martin[1, a]

[1] *Department of Organic Chemistry, Weizmann Institute of Science, 76100 Reḥovot, Israel*

a) Corresponding author: gershom@weizmann.ac.il

**Abstract.** Benchmarks that span a broad swath of chemical space, such as GMTKN55, are very useful for assessing progress in the quest for more universal DFT functionals. We find that the WTMAD2 metrics for a great number of functionals show a clear "Jacob's Ladder hierarchy"; that the "combinatorial" development strategy of Head-Gordon and coworkers generates "best on rung" performers; that the quality of the nonlocal dispersion correction becomes more important as functionals become more accurate for nondispersion properties; that fitting against small, unrepresentative benchmark sets leads to underperforming functionals; and that ωB97M(2) is currently the best DFT functional of any kind, but that revDSD-D4 functionals are able to reach similar performance using fewer parameters, and that revDOD-D4 in addition permits reduced-scaling algorithms. If one seeks a range-separated hybrid (RSH) GGA that also performs well for optical excitation energies, CAM-QTP-01 may be a viable option. The D4 dispersion model, with its partial charge dependence, appears to be clearly superior to D3BJ and even possibly NL. Should one require a double hybrid without dispersion model, noDispSD-SCAN is a viable option. Performance for the MOBH35 transition metal benchmark is different: the best double hybrids are competitive but not superior to ωB97M-V, which offers the best performance compromise for mixed main group-transition metal problems.

## INTRODUCTION

Recently, Goerigk and coworkers released the GMTKN55 benchmark[1] for density functional methods, which is both large (1,500 reaction energies) and chemically diverse (55 different types of properties, which can be grouped into five major categories: small molecule thermochemistry, large molecule reactions, barrier heights, intermolecular noncovalent interactions, and intramolecular/conformational ones). These authors covered a great many exchange-correlation functionals, but a number of more recent ones such as ωB97M(2)[2] went unexplored for technical reasons. In addition, a set of revised[3] DSD double hybrids has meanwhile been fitted to GMTKN55, improving substantially on the original ones.[4] The purpose of the present short paper is to summarize these data and stress the strengths and weaknesses of these methods.

In Perdew's "Jacob's Ladder" metaphor,[5] we climb a ladder from Hartree's "vale of tears" to the heaven of chemical accuracy by introducing an XC (exchange-correlation) functional and adding explicit dependence on additional pieces of information on each rung: on rung 1 (LDA) just the local density $\rho$; on rung 2 (GGA) the reduced density gradient $\rho^{-4/3}\nabla\rho$; on rung 3 (mGGA) the Laplacian or (contains equivalent information) the kinetic energy density $\tau$; on rung 4 the occupied orbitals (of which hybrids are the most common special case);[6] and finally on rung 5 the unoccupied orbitals (e.g., double hybrids[7] based on GLPT2, 2nd-order Görling-Levy[8] perturbation theory).

In the development of DFT functionals, physicists have traditionally strived for nonempirical functionals with parameters derived from known constraints, while chemists have embraced empiricism. The proliferation of empirical parameters in functionals like M06[9] and MN15[10] has inspired the "combinatorial optimization" approach of Mardirossian and Head-Gordon,[11] in which the parameter space is carefully pruned for statistical significance: this has resulted in the ωB97X-V,[11] B97M-V,[12] ωB97M-V,[13] and ωB97M(2) functionals.[2] From a different perspective, the DSD functionals (dispersion-corrected, spin-component scaled, double hybrid) developed in our group[3,4,14] reduce empirical parameters to a mere handful at the expense of introducing the GLPT2 step.

## COMPUTATIONAL METHODS

All quantum chemical calculations were performed using Q-CHEM[15] versions 5.1.1 and 5.2 running on the Faculty of Chemistry HPC facility. SG-3 or better integration grids were used throughout; basis sets were def2-QZVPP except for some anionic subsets where diffuse functions could matter much, where we employed def2-QZVPPD instead. For further details see Ref. [3]

The primary metric and "objective function" employed is the WTMAD2 (weighted mean absolute deviation, type 2) as defined by Goerigk et al, defined as follows:

$$\text{WTMAD2} = \frac{1}{\sum_i^{55} N_i} \cdot \sum_i^{55} N_i \cdot \frac{56.84 \text{ kcal/mol}}{|\overline{\Delta E}|_i} \cdot \text{MAD}_i$$

in which division by $|\overline{\Delta E}|_i$, the mean absolute value of all the reference energies for subset $i$, seeks to compensate for the different energy scales. $N_i$ the number of data points in the subset, and MAD$_i$ represents the mean absolute difference between calculated and reference reaction energies for subset $i$. We note that MAD is a more "robust statistic"[16] than the root mean square deviation, in the statistical sense that MAD is less sensitive to "outliers" than the RMSD (root mean square difference), conversely, RMSD is more useful for spotting outliers.

Aside from ωB97M(2) and the original[4] and revised[3] double-hybrid functionals, in order to "compare apples to apples", we also evaluated WTMAD2 in the exact same manner for some additional functionals on rungs 2–4. They were the GGAs PBE,[17] revPBE,[18] and B97-D3BJ;[19] the meta-GGAs TPSS,[20] revTPSS,[21] SCAN,[22] and B97M-V,[12] the global hybrid GGAs B3LYP,[23,24] PBE0,[25] and its analog revPBE0;[25] the global hybrid meta-GGAs PW6B95,[26] M06,[27] revM06,[28] M06-2X,[27] SCAN0,[29] and MN15,[10] the range-separated hybrids M11,[30] revM11,[31] CAM-QTP00 and CAM-QTP01,[32] as well as ωB97X-V,[11] and the range-separated hybrid meta-GGA ωB97M-V.[13] For additional references, see Refs. [1,3,33]

Parameters for the D3BJ[34] and D4[35] dispersion corrections were taken from Ref.[1] where necessary; where unavailable, they were obtained in this work by minimizing the GMTKN55 WTMAD2 in function of the dispersion terms using an adaptation of Powell's BOBYQA derivative-free bounded optimizer. [36]

## RESULTS AND DISCUSSION

Values of the WTMAD2 metric, as well as of its subtotals for the five principal subcategories, are given in Table 1 relative to the popular M06-2X functional, which is popular among mechanistic researchers owing to its good performance for both thermochemistry and barrier heights.

As concerns Jacob's Ladder, we can state that it is well preserved along the sequence LDA → PBE → SCAN → SCAN0 → SCAN0-2, and more clearly still along the pathway LDA → B97-D3 → B97M-V → ωB97M-V → ωB97M(2): each of these latter functionals is "best in class" on its rung except for B97-D3 which is slightly surpassed by revPBE-D3BJ.

Concerning the PBE variants, we note that PBEsol,[37] the solids version of PBE, leads to an unacceptable deterioration for molecules; Weitao Yang's modified revPBE,[18] in contrast, yields clearly improved performance. RPBE is comparable to PBE here, with some variation in the subsets.

The CAM-QTP01 RSH originated as a reparametrization of CAM-B3LYP[38] by the Bartlett group[32] in order to minimize deviations from the ionization potential theorem and thus improve performance for electronic spectroscopy (in what could be called a "globalized" take on the tuned range separated hybrids of Baer and Kronik,[39,40]). After fitting a D3BJ dispersion correction for it to the whole GMTKN55 set, we found that CAM-QTP01-D3BJ performs comparably to popular hybrids like PBE0-D3BJ and B3LYP-D3BJ for equilibrium thermochemistry — but, intriguingly, CAM-B3LYP-D3BJ beats them both, owing primarily to the noncovalent subsets.

As for empirical vs. nonempirical functionals: on rung 2 revPBE-D3BJ bests B97-D3BJ (principally for large-molecule reactions), but on rung 3, B97M-V[12] has a clear advantage over SCAN-D3BJ and indeed outperforms the two most popular *hybrids*, B3LYP-D3BJ and PBE0-D3BJ. On rung 4, the advance of ωB97M-V over revPBE0-D3BJ (the best nonempirical performer) is even more lopsided, as is the advantage on rung 5 of DSD-SCAN-D3BJ over the nonempirical SCAN0-2. [29] The superiority of empirical over "nonempirical" double hybrids[41] like SCAN0-2 was demonstrated in great detail by Goerigk and coworkers.[42]

**TABLE 1**. Performance of functionals for the GMTKN55 benchmark and its five major subsets, relative to the popular M06-2X functional, plus MAD (kcal/mol) for the MOBH35 transition metal barrier benchmark of Iron and Janes.[33] Blue shading represents better performance, and red shading worse performance, than M06-2X.

| Functionals | Rung | Empiricity | WTMAD2 | Thermo-chem. | Barrier heights | Large mol. reactions | Conformer Equilibria | Inter-molecular | MOBH35 Ref.[33] |
|---|---|---|---|---|---|---|---|---|---|
| LDA(SPW92) | 1 | 0 | 4.74 | 5.19 | 8.40 | 2.75 | 4.15 | 5.35 | 7.10 |
| PBEsol-D3BJ | 2 | 0[37] | 2.98 | 3.39 | 6.45 | 2.09 | 2.55 | 2.52 | |
| PBE-D3BJ | 2 | 0[17] | 2.18 | 2.44 | 4.99 | 1.86 | 1.54 | 1.79 | 5.00 |
| rPBE-D3BJ | 2 | 0[43] | 2.18 | 2.40 | 4.76 | 1.50 | 1.28 | 2.53 | |
| TPSS-D3BJ | 3 | 0[20] | 1.91 | 2.15 | 4.43 | 1.88 | 1.31 | 1.34 | 4.70 |
| B97-D3BJ | 2 | M[19] | 1.80 | 2.12 | 3.60 | 2.11 | 1.26 | 1.08 | 5.60 |
| revTPSS-D3BJ | 3 | 0[21] | 1.76 | 2.26 | 4.22 | 1.64 | 1.08 | 1.18 | 4.70 |
| revPBE-D3BJ | 2 | 0[18] | 1.74 | 2.07 | 4.24 | 1.58 | 1.26 | 1.11 | 5.30 |
| SCAN-D3BJ | 3 | 0[44] | 1.66 | 1.95 | 4.05 | 1.21 | 1.06 | 1.51 | 4.00 |
| M06-D3zero | 4MX | L[27] | 1.62 | 1.34 | 1.33 | 1.35 | 2.43 | 1.34 | 3.90 |
| SCAN0 | 4MX | 0[29] | 1.61 | 1.91 | 2.14 | 1.22 | 1.39 | 1.74 | 2.00 |
| CAM-QTP-02-D3BJ[a] | 4GR | S[45] | 1.52 | 1.53 | 1.90 | 1.18 | 1.56 | 1.65 | 2.90§ |
| CAM-QTP-01-D3BJ[b] | 4GR | S[32] | 1.42 | 1.46 | 1.94 | 1.12 | 1.48 | 1.40 | 2.40§ |
| CAM-QTP-00-D3BJ[c] | 4GR | S[32] | 1.35 | 1.92 | 2.23 | 1.18 | 1.04 | 1.05 | 4.10§ |
| PBE0-D3BJ | 4GX | 0[25] | 1.37 | 1.60 | 2.50 | 1.27 | 1.03 | 1.17 | 2.70 |
| B3LYP-D3BJ | 4GX | S[23,24] | 1.36 | 1.53 | 2.37 | 1.54 | 0.94 | 1.08 | 3.90 |
| M11 | 4MR | L[30] | 1.34 | 1.12 | 1.18 | 1.02 | 2.07 | 1.09 | 2.60 |
| B97M-V | 3 | M[12] | 1.33 | 1.39 | 2.08 | 1.45 | 1.44 | 0.74 | 3.00 |
| SCAN0-D3BJ[d] | 4MX | S[29] | 1.30 | 1.94 | 2.26 | 1.09 | 0.86 | 1.09 | 2.20§ |
| MN15-D3BJ | 4MX | L[10] | 1.21 | 1.16 | 1.08 | 0.94 | 1.74 | 0.97 | 2.50 |
| revM11 | 4MR | L | 1.20 | 1.31 | 1.58 | 1.19 | 1.32 | 0.83 | 2.10 |
| PW6B95-D3BJ | 4MX | S[26] | 1.15 | 1.22 | 1.64 | 1.38 | 1.07 | 0.75 | 2.50 |
| LC-wPBEh-D3BJ[e] | 4GR | S[46] | 1.15 | 1.54 | 1.97 | 1.15 | 0.93 | 0.74 | 2.50§ |
| revPBE0-D3BJ | 4GX | 0[25] | 1.13 | 1.59 | 2.00 | 0.98 | 0.93 | 0.80 | 2.90§ |
| BHandHLYP | 4GX | 0[47] | 1.16 | 1.84 | 1.70 | 1.17 | 0.79 | 0.80 | 3.50§ |
| CAM-B3LYP-D3BJ | 4GR | S[38] | 1.11 | 1.32 | 1.83 | 1.17 | 1.01 | 0.71 | 2.40§ |
| revM06 | 4MX | L | 1.11 | 1.18 | 1.09 | 1.07 | 1.37 | 0.82 | 2.10 |
| M06-2X | 4MX | L[27] | 1 by def. | 1 (unity) by definition | | | | | 2.90 |
| Actual values for M06-2X (kcal/mol) | | | **4.79** | **0.86** | **0.48** | **1.08** | **1.22** | **1.14** | |
| SCAN0-2 | 5 | 0[29] | 0.98 | 1.14 | 1.36 | 1.00 | 0.77 | 0.90 | 3.20 |
| ωB97X-D3BJ | 4GR | M[48] | 0.98 | 1.14 | 1.36 | 1.00 | 0.77 | 0.90 | 2.30 |
| ωB97X-V | 4GR | M[11] | 0.83 | 1.19 | 1.16 | 0.99 | 0.60 | 0.51 | 2.00 |
| ωB97M-D3BJ | 4MR | M[49] | 0.79 | 0.86 | 0.85 | 0.76 | 0.73 | 0.79 | 1.90 |
| ωB97M-V | 4MR | M[13] | 0.69 | 0.85 | 0.94 | 0.59 | 0.73 | 0.50 | 1.70 |
| B2GP-PLYP-D3BJ | 5 | S[50] | 0.67 | 0.73 | 0.86 | 0.61 | 0.52 | 0.74 | 2.60 |
| DSD-PBE-D3BJ | 5 | S[4] | 0.66 | 0.77 | 0.85 | 0.50 | 0.60 | 0.72 | 2.80 |
| DSD-PBEP86-D3BJ | 5 | S[4,14] | 0.65 | 0.64 | 0.93 | 0.45 | 0.53 | 0.85 | 2.70 |
| noDispSD-SCAN69 | 5 | S[3] | 0.62 | 0.68 | 1.09 | 0.62 | 0.39 | 0.64 | 3.10 |
| ωB97X-2(TQ) | 5R | M | 0.62 | 0.69 | 0.76 | 0.55 | 0.41 | 0.81 | 2.80 |
| revωB97X-2 | 5R | M | 0.59 | 0.68 | 0.77 | 0.54 | 0.41 | 0.67 | 2.70 |
| DSD-BLYP-D4 | 5 | S[4,35] | 0.59 | 0.68 | 0.79 | 0.55 | 0.55 | 0.52 | 2.80 |
| revDSD-PBEB95-D4 | 5 | S[3] | 0.56 | 0.74 | 0.64 | 0.41 | 0.64 | 0.46 | 2.00 |
| DSD-PBEP86-D4 | 5 | S[4,35] | 0.55 | 0.63 | 0.76 | 0.59 | 0.45 | 0.49 | 2.40 |
| DSD-PBEP86-NL | 6 | S[4,48] | 0.55 | 0.68 | 0.82 | 0.53 | 0.44 | 0.49 | 2.60 |
| DSD-PBE-D4 | 5 | S[4,35] | 0.55 | 0.71 | 0.80 | 0.52 | 0.43 | 0.48 | 2.40§ |
| DSD-SCAN-D4 | 5 | S[3] | 0.55 | 0.70 | 0.82 | 0.58 | 0.37 | 0.49 | 2.00 |
| revDSD-BLYP-D4 | 5 | S[3] | 0.54 | 0.66 | 0.71 | 0.53 | 0.39 | 0.55 | 2.10 |
| revDSD-PBEP86-D3BJ | 5 | S[3] | 0.51 | 0.62 | 0.64 | 0.51 | 0.37 | 0.50 | 2.10 |
| revDSD-PBE-D4 | 5 | S[3] | 0.51 | 0.75 | 0.73 | 0.49 | 0.35 | 0.43 | 2.10 |
| revDSD-PBEP86-NL | 5 | S[3] | 0.51 | 0.64 | 0.62 | 0.51 | 0.38 | 0.50 | [1.90] |
| revDOD-PBEP86-D4 | 5 | S[3] | 0.49 | 0.69 | 0.62 | 0.55 | 0.33 | 0.41 | 1.80 |
| revDSD-PBEP86-D4 | 5 | S[3] | 0.49 | 0.65 | 0.65 | 0.53 | 0.33 | 0.42 | 1.90 |
| ωB97M(2) | 5 | M[2] | 0.46 | 0.52 | 0.54 | 0.39 | 0.47 | 0.43 | 2.40 |

Empiricity: 0=nonempirical, S=small (≤6 empirical parameters), M=medium (about a dozen parameters), L=large (several dozen parameters)
Rung 4 subsets: G=GGA, M=mGGA, X=global hybrid, R=range-separated hybrid
 (a) $a_1$=0.270; $a_2$=7.338; $s_6$=1; $s_8$=1.037 (this work); (b) $a_1$=0.270; $a_2$=7.040; $s_6$=1; $s_8$=1.230 (this work); (c) $a_1$=0.377; $a_2$=6.186; $s_6$=1; $s_8$=2.415 (this work); (d) $a_1$=0; $a_2$=7.9042; $s_6$=1; $s_8$=0;[3]  (e) $a_1$=0.426; $a_2$=4.982; $s_6$=1; $s_8$=1.752 (this work)
 § M. A. Iron, personal communication (supplemental to Ref.[33])

At first sight, the DSD double hybrids offer only ωB97M-V like performance. However, some of that is due to deficiencies of the D3BJ dispersion model: when substituting the newer D4 model or the VV10 nonlocal dispersion functional (NL) as a "drop-in replacement", statistics improve considerably. Indeed, when the functional is reoptimized together with D4 over the whole GMTKN55 dataset, we are able to lower WTMAD2 for this revDSD-PBEP86-D4 functional to a value statistically equivalent to the "best in class" ωB97M(2). What's more, revDSD-PBEP86-D4 contains just four true fitting parameters, plus three additional parameters that can be fixed at reasonable values over a broad range — about one-third or one-half the parameters in ωB97M(2), depending on whether includes the three "semi-arbitrary" ones in the count.

Let us briefly turn to the major subsets. Barrier heights are the most demanding, with most functionals above M06-2X in the table running 2-5 times worse: the three main exceptions, M11, M06-D3, and MN15-D3BJ all show severely degraded performance for conformers (which are driven primarily by intramolecular interactions). If one wishes to exceed the already good performance of M06-2X for barriers *and* for thermochemistry, one needs to resort to either ωB97M-{V,D3BJ} or to a double hybrid, the latter with improved performance for other properties in the bargain.

(Speaking of subsets, we note that ωB97M-D3BJ is noticeably inferior to wB97M-V for noncovalent interactions as well, as is ωB97X-D3BJ relative to ωB97X-V. On the other hand, barrier heights and general thermochemistry are not impacted thus, and hence ωB97X-D3BJ and ωB97M-D3BJ become attractive for mechanistic exploration, as analytical 1$^{st}$ and 2$^{nd}$ derivatives are readily available.)

If one wishes to eliminate the dispersion correction entirely, the best option appears to be the noDispSD-SCAN double hybrid.[3] While somewhat inferior in performance to the revDSD-PBEP86 variants, it still represents a significant improvement over ωB97M-V, and moreover entails just three parameters (namely, $c_{C,DFT}$, $c_{2ab}$, and $c_{2ss}$).

How do these observations for the main group square with performance for transition metal systems? The final column of Table 1 shows MSDs for the MOBH35 (metal-organic barrier heights) of Iron and Janes.[33] One can see there that what holds true for main group does not necessarily hold true for transition metals: for instance, PBE0-D3BJ is clearly superior to B3LYP-D3BJ for transition metal reactions, even though the two functionals are comparable in quality for the main group. SCAN0 and SCAN0-D3BJ, despite unremarkable performance for main-group problems, are among the best performers for MOBH35. The three QTP range-separated hybrids CAM-QTP0$x$-D3BJ ($x$=0,1,2) perform comparably for main-group elements, but QTP01 is markedly superior for transition metals, and indeed the original CAM-B3LYP matches QTP01 for transition metals while outperforming it for GMTKN55. What CAM-QTP01 and CAM-B3LYP have in common with each other and with wB97X-V and wB97M-V are a relatively modest percentages of short-range HF exchange (23, 19, 17, and.15%, respectively), where CAM-QTP00 has 54% and CAM-QTP02 28%. We note that the Becke half-and-half hybrid (which has 50% over the whole distance range) likewise does not perform well for MOBH35 —and neither does B3LYP, which has 20% across the distance range (and hence does not benefit from reduced self-interaction error like the RSHes do). Double hybrids do not necessarily offer advantages for transition metal systems: In fact, the main group "top scorer" ωB97M(2) turns out to be inferior to ωB97M-V for MOBH35. (We note in passing that BH&HLYP actually has a lower WTMAD2 than B3LYP, owing to improved performance for barriers at the expense of small-molecule thermochemistry: handling both well in a global hybrid requires both elevated HF exchange and a meta-GGA, as was demonstrated first with the BMK functional,[51] and later applied to good effect in M06, M06-2X, and other Minnesota functionals. We also note in passing that revM06 represents a substantial improvement over the original M06 for both GMTKN55 and MOBH35.)

While our original DSD double hybrids, fitted to a smallish training set, leave something to be desired for MOBH35, the revised fits[3] to GMTKN55 like revDSD-PBEP86-D4 are comparable to the best performers, despite no transition metal systems at all being present in GMTKN55. This implies that refitting to the much larger and more diverse main-group dataset helped up capture broader molecular physics, rather than add to the plethora of functionals good for specific properties of specific system.

Results for the revDOD-PBEP86-D4 functional, which eliminates the same-spin GLPT2 term, are essentially indistinguishable for main-group systems and even slightly better for MOBH35. This is quite useful, since spin-opposite MP2 viz GLPT2 can be performed using reduced-scaling $O(N^4)$[52–54] or even $O(N^3)$ algorithms.[55]

A summary of our main conclusions has been given in the abstract. If we had to single out just two take-home messages, however, it would be that: (a) ωB97M-V is hard to beat as a "workhorse functional"; (b) for any further improvement beyond that, mildly empirical double hybrids such as revDSD-PBEP86-D4 look most promising.

## ACKNOWLEDGMENTS

This research was supported by the Israel Science Foundation (grant 1358/15), the Minerva Foundation, and the Helen and Martin Kimmel Center for Molecular Design (Weizmann Institute of Science). We thank Nitai Sylvetsky for critical reading of the manuscript and Dr. Mark Iron for helpful discussions and some additional data not present in the preprint version of Ref.[33]